\documentclass[a4paper,aip,reprint]{revtex4-1}
\usepackage{graphicx}
\usepackage{amsmath}
\usepackage{fancyhdr}
\usepackage{cancel}
\usepackage{color}
\setcitestyle{round}

\begin{document}
\title{Maximum Caliber: a general variational principle for dynamical systems} 
\author{Purushottam D. Dixit}
\affiliation{Department of Systems Biolog, Columbia University, New York, New York 10032}
\author{Jason Wagoner}
\affiliation{Laufer Center for Quantitative Biology, Stony Brook University, Stony Brook, NY, 11794}
\author{Corey Weistuch}
\affiliation{Department of Applied Mathematics and Statistics, Stony Brook University, Stony Brook, New York 11794}
\author{Steve Press\'e}
\affiliation{Deptartment of Physics and School of Molecular Sciences, Arizona State University, Tempe, AZ, 85281}
\author{Kingshuk Ghosh}
\affiliation{Department of Physics and Astronomy, University of Denver, Denver, Colorado 80208}
\author{Ken A. Dill}
\affiliation{Laufer Center for Quantitative Biology, Stony Brook University, Stony Brook, NY, 11794}
\affiliation{Department of Chemistry, Stony Brook University, Stony Brook, NY, 11794}
\affiliation{Department of Physics and Astronomy, Stony Brook University, Stony Brook, NY, 11794}

\begin{abstract}
We review here {\it Maximum Caliber} (Max Cal), a general variational principle for inferring distributions of paths in dynamical processes and networks.  Max Cal is to dynamical trajectories what the principle of {\it Maximum Entropy} (Max Ent) is to equilibrium states or stationary populations.  In Max Cal, you maximize a path entropy over all possible pathways, subject to dynamical constraints, in order to predict  relative path weights.  Many well-known relationships of Non-Equilibrium Statistical Physics -- such as the Green-Kubo fluctuation-dissipation relations, Onsager's reciprocal relations, and Prigogine's Minimum Entropy Production -- are limited to near-equilibrium processes.  Max Cal is more general.  While it can readily derive these results under those limits, Max Cal is also applicable far from equilibrium.  We give recent examples of MaxCal as a method of inference about trajectory distributions from limited data, finding reaction coordinates in bio-molecular simulations, and modeling the complex dynamics of non-thermal systems such as gene regulatory networks or the collective firing of neurons.  We also survey its basis in principle, and some limitations. 
\end{abstract}

\date{\today}

\maketitle

\section{Introduction: Non-equilibrium statistical physics: history and background}

We review here Maximum Caliber (Max Cal), a principle for inferring stochastic dynamics.  As an example of the problems being considered, consider vehicular traffic along a network of roads.  Suppose you know only a few average rates, such as the average number of cars traveling from  Chicago to New York per year.  MaxCal gives a way to use that limited information to estimate probabilities of all possible flows along individual roads.  As a principle, Max Cal resembles that of {\it Maximum Entropy} (Max Ent) used in equilibrium statistical mechanics for predicting the properties of materials, and also for inferring probabilities, in general, of states based on incomplete information, often used in model building.  

There has long been interest in establishing a variational principle for interpreting the statistical properties of dynamical systems~\citep{deGroot, Kreuzer, Lavenda, Kondepudi}.  Just as equilibrium statistical mechanics was formulated around a variational principle of entropy, a seemingly natural approach for nonequilibrium systems was to formulate corresponding variational principles of {\it entropy production} or dissipation. The most successful of these -- such as Prigogine's principle of minimum entropy production~\citep{prigogine1998modern} and Onsager's principle of least dissipation~\citep{onsager1953} -- are limited to near-equilibrium processes.  For example, the minimum entropy production principle (Min EPP) states that the stationary state distribution of a system interacting with multiple baths is the one that minimizes the total entropy production~\citep{klein1954principle}.  These near-equilibrium principles provide the quantitative underpinnings for continuum theories of flows of heat, particles, electrical currents and other conserved quantities.  This constitutes today's field of nonequilibrium thermodynamics, as it is now expressed in standard textbooks~\citep{Bird}.  However, `closeness to equilibrium' is often not well defined. Consequently, the range of application of the theoretical development to model experiments is unclear. 

Efforts to generalize these near-equilibrium principles to far-from-equilibrium situations have been largely unsuccessful. Nor has there been any effort to develop a probabilistic formalism going beyond properties of mean flows described by these near-equilibrium principles. However, starting in 1990s, there has been tremendous success in identifying principles in the form of \emph{fluctuation theorems} (FT). Briefly, fluctuation theorems relate the ratio of probability of observing a given trajectory (say $\Gamma$) and its time-reversed counterpart $\Gamma'$ to a dissipation-based quantity, such as the total entropy production during the trajectory~\citep{Evans:1993ix,Evans:1994tf,Gallavotti:1995vc,Gallavotti:1995gy,Lebowitz:1999tv,Maes:1998kn,Seifert:2005fu,Seifert:2012es}. The Crooks' FT~\citep{Crooks:1998wk} and its corollary the Jarzynski relation~\citep{Jarzynski:1997uj} are notable in that they pertain to work distributions. These relationships have extensive practical use, for example in constructing free energy profiles in single molecule pulling experiments~\citep{hummer2001free} and in clarifying the relationship between thermodynamics and information~\citep{Parrondo:2015cv,Sagawa:2012gx,Mandal:2012im}. Good reviews of these relationships are given in Refs.~\citep{Jarzynski:2011hl,Bustamante:2005im}.

The question we address in this review is whether there exists a variational principle for predicting distributions in dynamical processes, as a counterpart to the principle of maximum entropy for predicting distributions in equilibrium situations.  We suggest that Maximum Caliber (Max Cal) can be such a variational principle for dynamics, although much remains to be done.

\section{Variational principles predict probability distributions in statistical physics}

\subsection{The Boltzmann Distribution for equilibria as an application of Max Ent}

The Second Law of Thermodynamics predicts that matter tends to alter its degrees of freedom to reach the maximum of {\it entropy} at equilibrium.  The Second Law applies on the macroscale to explain how heat exchanges from hot to cold bodies, particles diffuse from crowded to low-density regions, or pressures tend to equalize.  On the microscale, the Second Law takes a statistical form, as first recognized by Maxwell and Boltzmann.

Specifically, it was argued that the most probable probability distribution of a system at equilibrium is the one that maximizes the entropy. This microscopic variational principle proposed by Boltzmann  was later generalized by Gibbs.  Briefly, the probability distribution $\{p_i\}$ over microscopic states $\{i \}$ of a macroscopic system, say in thermodynamic equilibrium with a surrounding heat bath, can be obtained by maximizing the entropy $S$,
\begin{eqnarray}
S = -\sum_i p_i \log p_i \label{eq:s0}
\end{eqnarray}
over $\{p_i\}$, subject to a constraint of average energy
\begin{eqnarray}
 \sum_i p_i E_i \label{eq:c1} =\bar E
\end{eqnarray}
where $E_i$ is the energy of microscopic state $i$, $\bar E$ is the average energy of the system, and $k_B T$ are Boltzmann's constant and temperature and  that probabilities are normalized quantities
\begin{eqnarray}
\sum_i p_i = 1. \label{eq:c2}
\end{eqnarray}
The resulting Gibbs-Boltzmann distribution after maximizing the entropy is
\begin{eqnarray}
p_i^\ast \propto e^{-E_i/k_BT},
\label{eq:gb0}
\end{eqnarray}
where $p_i^\ast$ are the probabilities that satisfy these conditions.  The Gibbs-Boltzmann distribution is at the heart of equilibrium statistical physics textbooks.  In short, given a model for the energies, $E_i$ of a system, and given a measured value of the first moment of the energy $\langle E \rangle = k_{B} T$ (which in this case means a known value of the temperature $T$), Eqn.~\ref{eq:gb0} predicts the probability distribution over the states $i = 1, 2, 3, \ldots$.

\subsection{Maximum Caliber: A variational principle for dynamical systems}

 As shown above, Max Ent appears as a principle of thermal material equilibrium.  However, as noted later in this review, Maximum Entropy is a rather more general principle of inference and model-making about any types of probability distributions.  It is not restricted to predictions of thermodynamics, materials, or equilibrium.  In particular, as noted below, it can also be used in dynamical modeling to infer distributions of path probabilities.  When used in dynamics, it has been called maximum caliber (Max Cal)~\citep{Jaynes80}. Max Cal is a trajectory-based method of dynamics~\cite{Filyukov2,Haken86,EricSmith, Cecille,PresseRMP2013,dixit2014inferring,dixit2015inferring,dixit2015stationary}.  It seeks the probabilities of paths or trajectories of individual particles, molecules or agents.  In short, the idea is to find the path distribution that maximizes a path entropy, subject to imposed constraints. 

Consider a system whose coordinates are collectively described by the variable $\sigma$. For simplicity, we assume that the system evolves in a discrete-time and discrete-state fashion. Let $\{ \Gamma\}$ be the set of all possible trajectories, individually given by $\Gamma = \{\sigma_{T_i}, \sigma_{T_i + 1} , \dots ,  \sigma_{T_f}\}$, that the system can take between time points $T_i$ and $T_f$. Finally, let $p_\Gamma$ be the probability distribution defined over the ensemble $\{ \Gamma \}$ of paths.

Let $F(\Gamma)$ be a functional defined on the space of paths. Examples of $F$ include the total flux of mass/heat carried by the path, the average dissipation along the path, or the average energy along the path. Analogous to the equilibrium problem, imagine a situation where we want to infer the distribution $p_\Gamma$ over the paths while constraining the average
\begin{eqnarray}
\langle F \rangle = \sum_{\Gamma} p_\Gamma F(\Gamma).  \label{eq:lfr}
\end{eqnarray} Note that there are potentially infinitely many  probability distributions $p_\Gamma$ that are consistent with such constraints. Analogous to the equilibrium situation, we maximize the entropy
\begin{eqnarray}
S = -\sum_\Gamma p_\Gamma \log \frac{p_\Gamma}{q_\Gamma}
\end{eqnarray} now defined as  a {\it distribution over paths}, subject to constraint in Eqn.~\ref{eq:lfr} and normalization. Here, $q_\Gamma$ is some reference/prior distribution over paths. An implicit assumption in the equilibrium maximum entropy methods is priors are the same for all micro states.

The constrained maximization problem is solved by introducing Lagrange multipliers. We write the unconstrained optimization function, popularly known as the Caliber $C$,
\begin{eqnarray}
C &=& -\sum_\Gamma p_\Gamma \log \frac{p_\Gamma}{q_\Gamma} - \gamma \left (\sum_{\Gamma} p_\Gamma F(\Gamma) - \langle F \rangle \right ) \nonumber \\ &+& \alpha  \left ( \sum_\Gamma p_\Gamma - 1\right ).\label{eq:caliber}
\end{eqnarray}

In Eq.~\ref{eq:caliber},  $\gamma$ is a Lagrange multiplier that tunes the ensemble average $\langle F \rangle$ and $\delta$ ensures normalization. After maximization, we find
\begin{eqnarray}
p_\Gamma = \frac{q_\Gamma e^{-\gamma F(\Gamma)}}{\mathcal Z} \label{eq:maxcal0}
\end{eqnarray}
where
\begin{eqnarray}
\mathcal Z = \sum_\Gamma q_\Gamma e^{-\gamma F(\Gamma)},
\label{eq:maxcal1}
\end{eqnarray}
a sum of weights over paths, is the dynamical equivalent of a partition function. Eqs.~\ref{eq:maxcal0} and  ~\ref{eq:maxcal1} are not particularly useful for computations as they stand.  They are rendered practical, for example, when the value of $\gamma$ is known and related to an average flux or rate by the derivative relationship,
\begin{eqnarray}
-\frac{\partial  }{\partial \gamma}\log \mathcal Z = \langle F \rangle.
\end{eqnarray}
While the expressions above follow from using constraints with no associated uncertainty (i.e. hard constraints), Ref.~\citep{PresseRMP2013} discusses generalizations of the results of this section
to problems involving constraints with associated uncertainty.

\section{The Max Cal principle generates several main results of nonequilibrium statistical physics}

\subsection{Max Cal modeling of non-equilibrium stationary states~\citep{HazoglouJCP2015}}

\begin{figure}
        \includegraphics[scale=0.25]{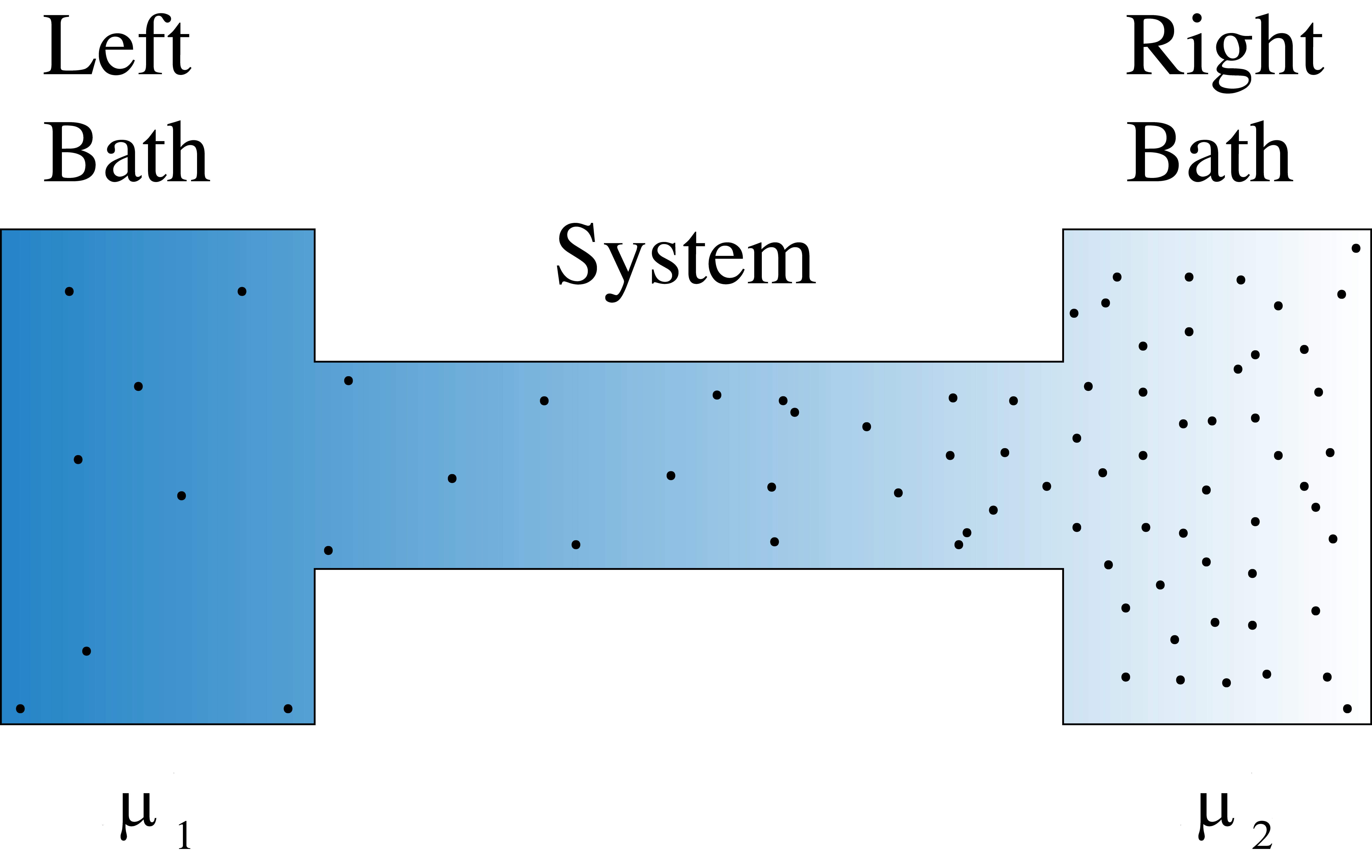}
        \caption{Consider a system connecting two large baths of particles. Imagine that the density of `stuff' (particles/heat) in the two baths is different leading to a constant flow from the right bath the the left bath.  Reproduced from The Journal of Chemical Physics 143, 051104 (2015), with the permission of AIP Publishing.\label{fg:flow}}
\end{figure}

 First consider the flow of particles between two baths (see Figure~\ref{fg:flow}). On the right is a bath with higher density of particles.  It is connected \emph{via} a small conduit (the `system') to a bath on the left having a lower density of particles.  After an initial period, the system reaches a steady state with a constant flux of particles from the right to left baths.

When the system is macroscopic, Fick's law of diffusion describes both the initial transient dynamics as well as the system's steady state flux.  When the number of particles is low, there are frequent violations of the net flux direction; particles may climb up the concentration gradient. However, no microscopic theory exists to quantify flux distributions. Here, we discuss the consequences of inferring Max Cal distributions over trajectories by constraining average flux quantities. Notably, the Max Cal-inferred distribution is consistent with many results in near-equilibrium thermodynamics of non-equilibrium stationary states.

Consider a system with time-dependent fluxes (mass, heat, etc.) (see Fig.~\ref{fg:flow}). We want to describe the distribution $p_\Gamma$ over the trajectories $\Gamma$ of this system. The ensemble average flux of some quantity $a$ (say heat or mass) over the ensemble of trajectories at a fixed time $t$ is given by
\begin{equation}
J_a \left( t\right) = \left \langle j_{a\Gamma} \left(t \right) \right \rangle 
= \sum_{\Gamma} p_{\Gamma}  j_{a\Gamma} \left(t \right) . 
\label{e:FluxAvg}
\end{equation}
In Eq.~\ref{e:FluxAvg}, $j_{a\Gamma}(t)$ is the flux of quantity $a$ at time $t$ in the trajectory $\Gamma$.

Consider the case where we want to infer the distribution $p_\Gamma$ over trajectories that is consistent with two macroscopic fluxes $J_a(t)$ and $J_b(t)$ where  $a$ and $b$ characterize two types of fluxes such as particles and heat or two types of particles. Potentially, there are infinitely many distributions $p_\Gamma$ that are consistent with these two constraints. We choose the one with the maximum path-entropy (caliber). We write down the caliber
\begin{eqnarray}
C &=& - \sum_{\Gamma} p_{\Gamma} \log \frac{p_{\Gamma} }{q_{\Gamma} } +  \sum_t \lambda_a \left( t \right) \left(\sum_{\Gamma} p_{\Gamma}  j_{a\Gamma} \left(t \right) -J_a \left( t\right)  \right) \nonumber \\ &+& \sum_t \lambda_b \left( t \right) \left(\sum_{\Gamma} p_{\Gamma}  j_{b\Gamma} \left(t \right) -J_b \left( t\right)  \right)  + \alpha \left( \sum_{\Gamma} p_{\Gamma} - 1\right). \nonumber \\
\label{e:CaliberConstraints}
\end{eqnarray}
In Eq.~\ref{e:CaliberConstraints},  $\lambda_a(t)$ and $\lambda_b(t)$ are time dependent Lagrange multipliers enforcing the constraint regarding known fluxes. Maximizing the Caliber gives:
\begin{eqnarray}
p_{\Gamma} = \frac{q_{\Gamma}}{Z} \ \mbox{exp} \ 
\left[ \lambda_a \left( t \right)  j_{a\Gamma}\left(t \right) +
 \lambda_b \left( t \right)  j_{b\Gamma} \left(t \right)\right] \label{eq:pGamma}\\
Z =  \sum_{\Gamma} q_{\Gamma}\ \mbox{exp} \ 
\left[ \lambda_a \left( t \right)  j_{a\Gamma}\left(t \right) +
 \lambda_b \left( t \right)  j_{b\Gamma} \left(t \right)\right]. 
\label{e:PartitionFn}
\end{eqnarray}
Note that at equilibrium we have $\lambda_{a,b}(t) = 0~\forall~t.$

Observable rate quantities, which are ensemble-averaged over pathways, can be obtained as derivatives of the dynamical distribution function:
\begin{eqnarray}
\frac{\partial \ \mbox{log} \ Z}{\partial \lambda_a \left( t \right) } &=& J_a \left( t \right) \\
\frac{\partial^2 \ \mbox{log} \ Z}{\partial \lambda_a \left( t \right) \partial \lambda_b \left( \tau
 \right) } &=&
\left \langle  j_{a\Gamma}\left(t \right)  j_{b\Gamma}\left(\tau \right)\right \rangle - 
\left \langle  j_{a\Gamma}\left(t \right)\right \rangle
 \left \langle  j_{b\Gamma}\left(\tau \right)\right \rangle. \nonumber \\
\label{e:firstsecondMoment}
\end{eqnarray}

We stress that the distribution predicted by Eq.~\ref{eq:pGamma} is a `model prediction' based on the constraint of average fluxes. This distribution does not account for other important physical characteristics of the system, for example, the amount of heat dissipation. Below, we show that this model allows us to capture many well-known near equilibrium results.  We also note its  limitations.

\subsection{The Green-Kubo relations from Max Cal}

Suppose the system described above is near equilibrium; \emph{i.e.}, the fluxes are small. We expand fluxes at some arbitrary time (say $t=0$) around Lagrange multipliers $\lambda(t) =0$. That is, we expand $\left \langle j_{a\Gamma} \left( t \right) \right \rangle = J_a \left(t \right)$ at $t=0$ to first order around $\lambda_a \left(\tau\right), \lambda_b \left(\tau\right)=0$ for all $\tau$ in the past:
\begin{eqnarray}
J_a \left(0 \right) & \approx & 
\sum_{\tau} 
\left[ \left. \frac{\partial \left \langle j_{a\Gamma} \left( 0 \right) \right \rangle  }{\partial \lambda_a \left(\tau\right)} \right |_{\lambda=0} \lambda_a \left(\tau\right) 
+ 
\left. \frac{\partial \left \langle j_{a\Gamma} \left( 0 \right) \right \rangle  }{\partial \lambda_b \left(\tau\right)} \right |_{\lambda=0} \lambda_b \left(\tau\right) \right] \nonumber \\
\end{eqnarray}

At steady state, the Lagrange multiplier do not depend on time. We have $\lambda_a(t) = \lambda_a~\forall~t.$ Thus,
\begin{eqnarray}
 J_a \left(0 \right) & \approx & 
 \lambda_a \sum_{\tau}  \left. \left \langle j_{a\Gamma} \left( 0 \right)  j_{a\Gamma} \left( \tau \right) \right \rangle \right |_{\lambda=0} \nonumber \\ &+&  \lambda_b \sum_{\tau}  \left. \left \langle j_{a\Gamma} \left( 0 \right) j_{b\Gamma} \left( \tau \right) \right \rangle \right |_{\lambda=0} \nonumber \\ \label{e:GreenKubo}
\end{eqnarray}
In Eq.~\ref{e:GreenKubo}, the $\lambda$'s can be interpreted as the driving forces. Moreover, the cross-correlations $\sum_{\tau}    \langle j_{a\Gamma} ( 0 )  j_{a\Gamma} ( \tau )  \rangle  |_{\lambda=0} $ quantify `flux' fluctuations at equilibrium and thus can be identified as the transport coefficients.  With this identification of terms, the result is just the Green-Kubo relationship~\cite{green1952,green1954,kubo1957}.

\subsection{Onsager's reciprocal relations from Max Cal}

Max Cal can also capture Onsager's reciprocal relationships. Onsager considered near-equilibrium systems in which fluxes are linearly proportional to the imposed forces~\cite{onsager1931a,onsager1931b}:
\begin{eqnarray}
J_a & = & L_{aa} \lambda_a + L_{ab} \lambda_b \\
J_b & = & L_{ba} \lambda_a + L_{bb} \lambda_b.
\label{e:Onsager}
\end{eqnarray}

Using  Eqns. \ref{e:firstsecondMoment}-\ref{e:GreenKubo}, we have
\begin{eqnarray}
L_{ab} = \displaystyle{ \left. \sum_\tau \frac{\partial^2 Z}{\partial \lambda_a(0)\partial \lambda_b(\tau)} \right |_{\lambda = 0}}.\label{e:OnsagerMax Cal}
\end{eqnarray}
At the same time, we have at $\lambda = 0$ (equilibrium),
\begin{eqnarray}
\sum_{\tau}  \left. \left \langle j_{a\Gamma} \left( 0 \right)   j_{b\Gamma} \left( \tau \right) \right \rangle \right |_{\lambda=0} &=& \sum_{\tau}  \left. \left \langle j_{a\Gamma} \left( \tau \right)   j_{b\Gamma} \left( 0 \right) \right \rangle \right |_{\lambda=0}\nonumber \\ &=& \displaystyle{ \left. \sum_{\tau} \frac{\partial^2 \ \mbox{log} \ Z}{\partial \lambda_b \left( 0 \right) \partial \lambda_a \left( \tau \right)} \right|_{\lambda=0}  = L_{ba}}. \nonumber \\ \label{e:reversible}
\end{eqnarray}
In Eq.~\ref{e:reversible}, we have assumed that both fluxes have the same parity under time reversal (symmetric or anti-symmetric) and invoked microscopic reversibility of trajectories at  equilibrium state. As a result, we have $L_{ab} = L_{ba}$ which is exactly Onsager's reciprocal relationship.

\subsection{Prigogine's Principle of Minimum Entropy Production from Max Cal}

An interpretation of Prigogine's principle of minimum entropy production is as follows. Consider a near-equilibrium system with two coupled flows. Imagine that one of the flows (flow of $a$) is driven by a force while the flow of $b$ is unconstrained. The flux of $b$ at steady state is predicted to be that which has the minimum rate of entropy production~\citep{seifert2008stochastic,tome2012entropy}. First, we consider the standard derivation of the principle. If $S$ is the state entropy, the rate $dS/dt$, of entropy production, in a system carrying two fluxes $J_a$ and $J_b$ is given by
\begin{equation}
\sigma = \frac{dS}{dt} = J_a \lambda_a + J_b \lambda_b
\end{equation}
where $\lambda_a$ and $\lambda_b$ are driving gradients. Now, near equilibrium, the Onsager relationships give
\begin{equation}
        \sigma=L_{aa} \lambda_a^2 +2L_{ab}\lambda_a \lambda_b+ L_{bb} \lambda_b^2.
\end{equation}

The minimal entropy production rate with respect to variations in $\lambda_b$ is given by,
\begin{equation}
        \frac{\partial \sigma}{\partial \lambda_b}=2(L_{ab} \lambda_a+ L_{bb} \lambda_b)=2J_b=0,
\end{equation}
which, correspondingly also predicts that $J_b = 0$ \cite{Kon1998}.

The same principle can also be derived from Max Cal.   First, we express the Caliber as
\begin{eqnarray}
        \mathcal{C} &=&-\sum_\Gamma p_\Gamma \ln \left( \frac{ p_\Gamma}{q_\Gamma} \right) \nonumber \\
        &=& \ln Z- \sum_{t} \left[ \lambda_a(t)  J_a(t) + \lambda_b(t) J_b(t) \right].
\end{eqnarray}

Maximizing $\mathcal{C}$ with respect to $\lambda_b$, 
\begin{equation}
        \begin{split}
                \frac{\partial \mathcal{C}}{\partial \lambda_b(\tau)}&=-\sum_{t} \left[ \lambda_a(t) \frac{\partial J_a(t)}{\partial \lambda_b(\tau)} + \lambda_b(t)\frac{\partial J_b(t)}{\partial \lambda_b(\tau)} \right] \\
&\approx -\lambda_a L_{ab}-\lambda_b L_{bb}+\mathcal{O}(\lambda^2) =-J_b =0.
        \end{split}
\end{equation}
Thus, the force-flux relationship derived using Max Cal is the same as the entropy-production argument above.  Notably,  Max Cal makes useful predictions beyond the linear regime that can be explicitly tested. The Caliber is maximized when
\begin{eqnarray}
\sum_{t} \left[ \lambda_a(t) \frac{\partial J_a(t)}{\partial \lambda_b(\tau)} + \lambda_b(t)\frac{\partial J_b(t)}{\partial \lambda_b(\tau)} \right]=0.\label{eq:nonlinearprig}
\end{eqnarray} 
So, given how $J_a$ and $J_b$ depend on the imposed thermodynamic gradients $\lambda_a$ and $\lambda_b$, solving Eq.~\eqref{eq:nonlinearprig} gives the gradient $\lambda_b$ to which the system adjusts itself when it is not constrained. 

 \subsection{Max Cal gives Fick's Law of diffusion, including the `few-molecule' limit}  
 
\begin{figure}
	\includegraphics[scale=0.5]{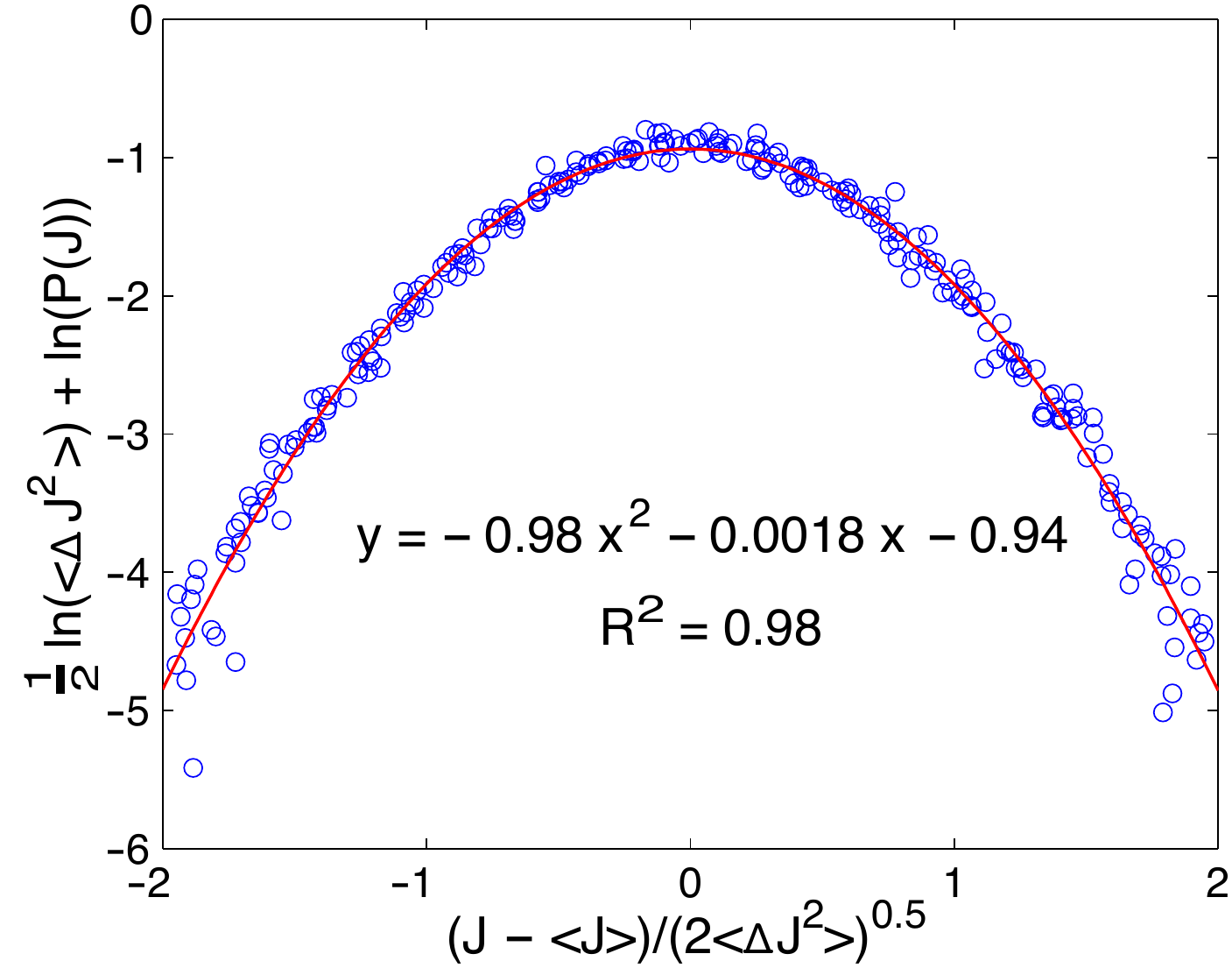}
	\caption{Max Cal predicts the distribution of microscopic fluxes to be Gaussian (red line). Experimental data shown in blue circles. See~\citep{Frosso} for details. Reprinted with permission from Effrosyni Seitaridou, Mandar M. Inamdar, Rob Phillips, et al., Journal of Physical Chemistry B. Copyright 2007 American Chemical Society.\label{fg:3c}}
\end{figure}

 The maximization of path entropy predicts diffusion and the Fokker Planck equation~\cite{WangASS2006, Haken86}.  Fick's law expresses that the macroscopic average rate of particle diffusion, driven by a gradient, is:
\begin{equation}
\langle J \rangle = - D \frac{d c}{d x}\end{equation}
where $D$ is the diffusion constant, $\langle J \rangle$ is the average flux and $dc/dx$ is the macroscopic gradient of particle concentration.  But, we are interested here in more microscopic detail.  What is the full rate distribution?  For example, what is the second moment of flux, $\langle J^2 \rangle - \langle J \rangle^2$?  This is a simple problem that can be solved in various ways, including using the Boltzmann transport equation~\cite{Huang}.  The difference is that Max Cal focuses on distributions of paths, rather than particle concentrations, and is not limited to near-equilibrium assumptions~\cite{Ghosh06,Frosso,PresseRMP2013}.  While we don't give the details here, we note that Max Cal gives the full rate distribution, which is predicted to be Gaussian, and verified experimentally (see Fig.~\ref{fg:3c})  The dynamical constraint that is imposed here is a hopping probability between discretized space which is equivalent to knowing the value of the diffusion constant $D$~\cite{Ghosh06,Frosso}.  Max Cal's trajectory-based approach predicts the flux distribution, including variances, which cannot be done using Fick's law or the diffusion equation or Boltzmann's transport equation, which are based on concentrations/density.  This Max Cal modeling gave the new result that Fick's Law holds even for gradients down to as small as a few molecules.  Closely related is a treatment that maximizes the path entropy while constraining the action ($A$) averaged over all paths~\cite{WangASS2006}.  Wang has also used that approach to derive other phenomenological laws such as Ohm's law and Fourier's law of heat flow~\cite{WangASS2006}. Earlier work\cite{Haken86} has shown that path entropy maximization can be used to derive generalized Fokker-Planck equation using first and second moments of coordinate variables -- other than action -- as constraints.

 \subsection{Markov models give the dynamics that maximize the Caliber for particular data}

 Dynamics is often modeled as a Markov process, where the probability of transition to a state depends only on the most recent history.  This is also the basis for Hidden Markov Models widely used in data analysis.  What is the justification for the Markov assumption?  It is found that Markovian dynamics uniquely maximize the Caliber, depending on the nature of the form of the measured rate data~\cite{Hao,JulianJCP2012}.  Depending on the data that is used as constraints, the probability over paths partitions into a product over transition probabilities depending on the state occupied at the previous time point~\cite{Filyukov1,Filyukov2,Stock1,GhoshJCP2011,Hao,JulianJCP2012}. For example if constraints are defined using average number of transitions $N_{i,j}$ between two consecutive time steps -- $i$ is the state at some instant of time and $j$ is the state at the next time step -- we recover traditional master equation~\cite{Hao} and Lagrange multipliers relate to reaction rates~\cite{Stock1}.  This also applies when the measured jump statistics include multiple time steps~\cite{JulianJCP2012} or memory~\cite{Stock2}.

\section{Max Cal infers dynamical distributions from limited information}

\subsection{Inferring the full rate matrix of a network from its state populations}

\begin{figure*}
	\includegraphics[scale=0.4]{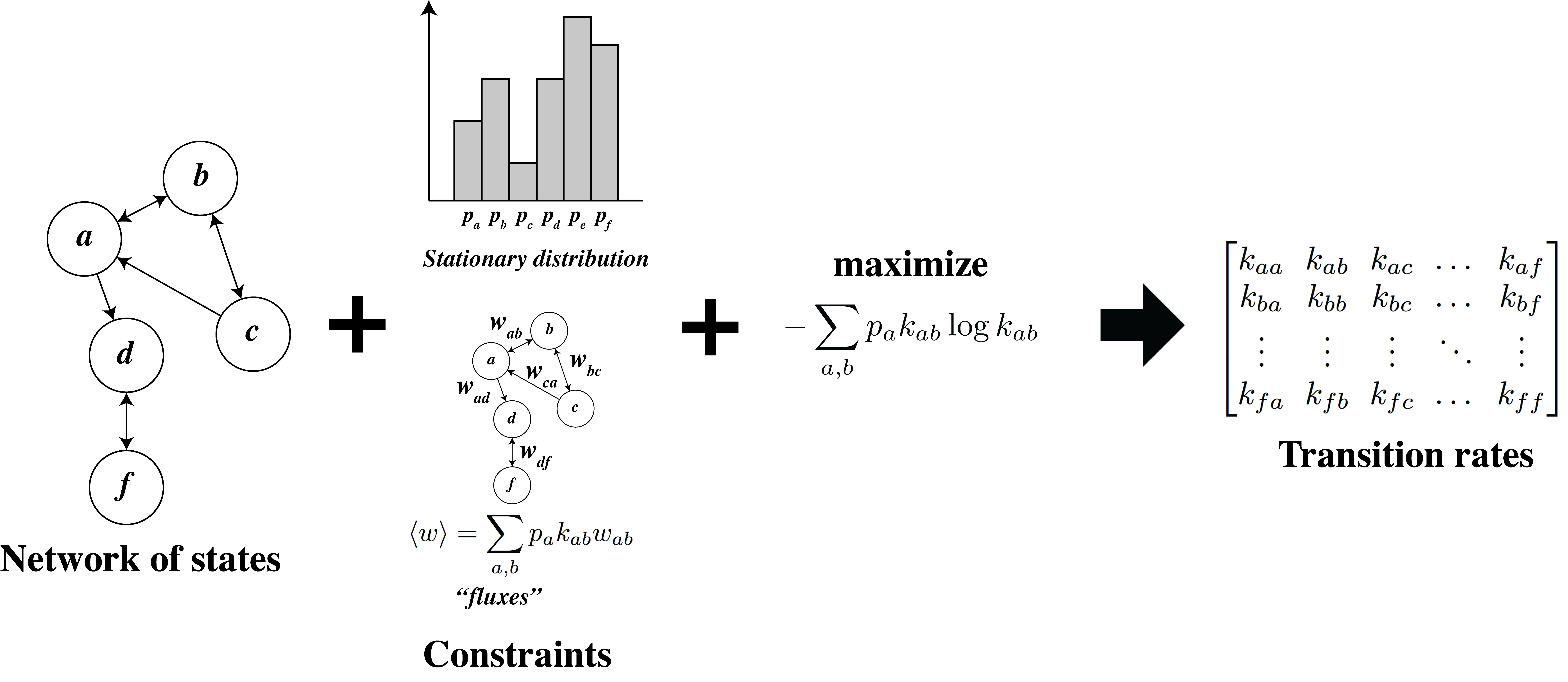}
	\caption{The Max Cal approach predicts the Markovian transition rates $k_{ab}$ over all routes between states $a$ and $b$, given only the steady-state populations at the nodes, in addition to an average global rate observable.}
	\label{fg:schematic_pdd_markov}
\end{figure*}

In single-molecule experiments and simulations of biomolecules, we may know or assume populations at the nodes of a network, while having very limited information on node-to-node jump dynamics. Consider a situation in which a single biomolecule visits a set of conformations or states $\{a, b, c, \dots \}$, including metastable states. The limited information can include the stationary distribution $\{p_a \}$, stationary state averages $\langle E \rangle = \sum p_a E_a$ over nodes, or path ensemble averages $\langle J \rangle = \sum_{a,b} p_a k_{ab} J_{ab}$, where $\{k_{ab} \}$ are the the transition probabilities (or equivalently, transition rates, when considering a continuous time Markov process) between these states. Our goal is to infer the set of transition rates $\{k_{ab}\}$ from infinitely many Markov processes that are consistent with such limited data.  What then is the best Markov model that we can infer? Recently, Dixit et al. used Max Cal to derive a functional form of the rate constants to reproduce a stationary distribution $\{p_a \}$ and dynamical path-based constraints  $\langle J \rangle$~\citep{dixit2014inferring,dixit2015inferring,dixit2015stationary} (see Fig~\ref{fg:schematic_pdd_markov}). They showed that the transition rates -- satisfying detailed balance --  are proportional to the square root of the stationary probability distribution $k_{ab} \propto \sqrt{p_b/p_a}$. This relationship is validated on the dynamics of small peptides and genetic networks~\citep{dixit2014inferring,dixit2015inferring,dixit2015stationary}. Wan et al. used Max Cal to modify the Markov state model describing the dynamics among metastable states of wild type peptides to capture the effect of mutations on folding dynamics~\citep{wan2016maximum} and Zhou et al. used it to study the effect of protein-protein interactions on transitions among the metastable states of proteins~\citep{zhou2017bridging}.

\subsection{Finding good reaction coordinates in molecular simulations}

When performing molecular simulations of chemical or physical processes, it is often challenging to learn the dominant reaction paths.  These are of interest because reaction paths are the essential coordinates which define chemical  `mechanisms'.   Many methods have been developed to find reaction paths~\citep{bolhuis2002transition}.  For example, metadynamics is an adaptive simulation technique to explore free energy landscapes along a few collective variables, such as reaction coordinates.  However, the challenges are to choose good collective variables, and to identify `slowly changing' collective variables that are suitable for sampling rare events, for example, in barrier crossing.   Based on the Max Cal method of Dixit et al., Tiwary et al. developed novel metadynamics-based algorithms for fast identification of reaction coordinates by maximizing timescale separation~\citep{tiwary2016spectral,tiwary2016wet,tiwary2017predicting,tiwary2017molecular}. In this way, Tiwary shed light on the molecular mechanisms orchestrating unbinding of streptavidin from the biotin-streptavidin complex~\citep{tiwary2017molecular}.

\subsection{Modeling networks that are biochemical or social}

\begin{figure*}
	\includegraphics[scale=0.5]{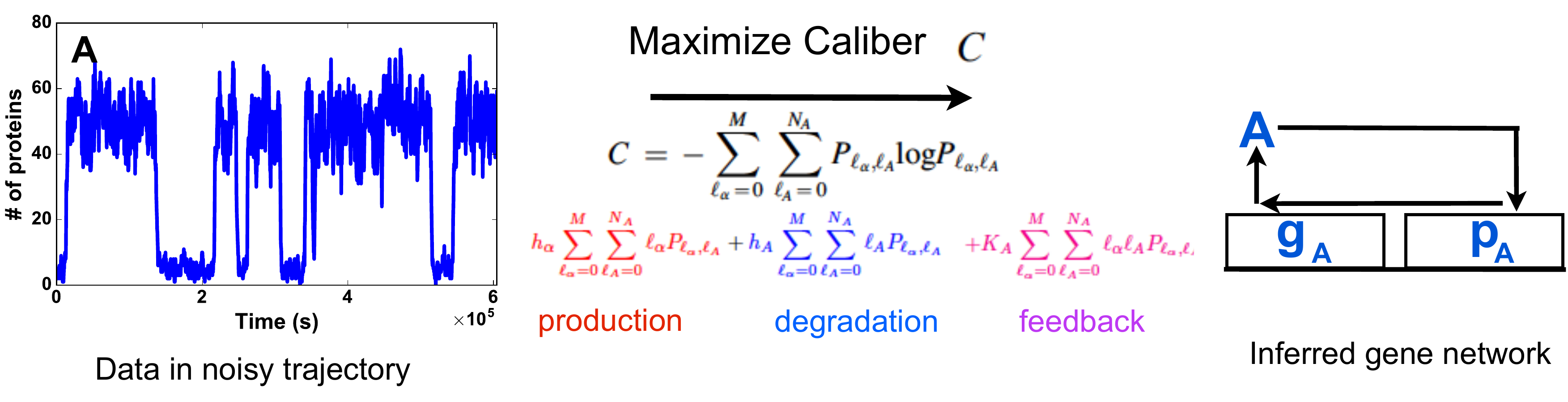}
	\caption{Max Cal can be used to infer details of the underlying gene network using experimentally measured noisy gene expression trajectory. Max Cal develops a model by maximizing the Caliber (${\cal C}$) constructed from the path entropy and three constraints of production, degradation and feedback for a single gene auto activating circuit. Details of the model and methodology can be found in~\cite{FirmanBJ2017}. Figure reprinted from Firman et al. (2017) with permission from Elsevier. \label{fig:infernetwork}}
\end{figure*}

Cells in an isogenic populations often have widely fluctuating protein copy numbers due to stochastic gene expression~\citep{PaulssonNature2004}.  The noisy time profiles of protein expression are the stochastic trajectories to which Max Cal can be applied to infer a predictive model (see Figure~\ref{fig:infernetwork}). These problems are ubiquitous in genetic networks and are particularly challenging when there is feedback. Feedback often involves interactions between multiple species not directly observable in experiments. Typically, only one or two types of proteins can be seen -- far too few compared with the actual number of molecular actors involved. These are examples of underdetermined problems with limited information. How do we infer microscopic parameters for these models? In normal, ``forward'', modeling in physics, a model is assumed and dynamical equations of motion are written. Predictions are then made and compared to the data. For the example of the toggle switch, you could start with master equations describing the dual-negative feedback loop and subsequently make predictions on the basis of this model. The challenge with this approach is that the model parameters are either introduced in an \emph{ad hoc} fashion or adjusted to fit the data. Thus predictions are sensitive to the choice of parameters. 

By contrast, in the inverse modeling approach, the goal is to learn a model from the data, with otherwise minimal assumptions.  In  this case, using Max Cal with observed particle-number fluctuations is sufficient to predict the dynamics, minimizing assumptions and adjustable parameters (see Figure~\ref{fig:infernetwork}). This was demonstrated in two synthetic gene circuits: i) in a positive feedback (PF) circuit in which a gene auto-activates itself~\cite{FirmanBJ2017} and ii) in genetic toggle switch (TS) in which two genes repress each other~\citep{SteveToggle,PresseRMP2013}.  The information input was:  1) protein synthesis; 2) protein turnover; and 3) effective coupling between species (positive feedback in case of PF and negative feedback in case of TS; see Figure~\ref{fig:infernetwork} for PF circuit). The success of the method was shown using synthetic time traces generated from known models (with seven parameters for PF and four parameters for TS) using a Gillespie simulation. Max Cal was found to capture the same qualitative and quantitive information with few Lagrange multipliers. Moreover, MaxCal inferred underlying rates accurately and produced an effective feedback parameter~\cite{FirmanBJ2017}. The MaxCal framework works directly on the trajectory space and is readily amenable to further including raw trajectories that are likely to be given in fluorescence (observed in typical experiments) instead of particle numbers~\cite{FirmanBJ2017}. This approach can help analyze raw noisy fluorescence trajectories instead of protein numbers.

\begin{figure}
	\includegraphics[scale=0.4]{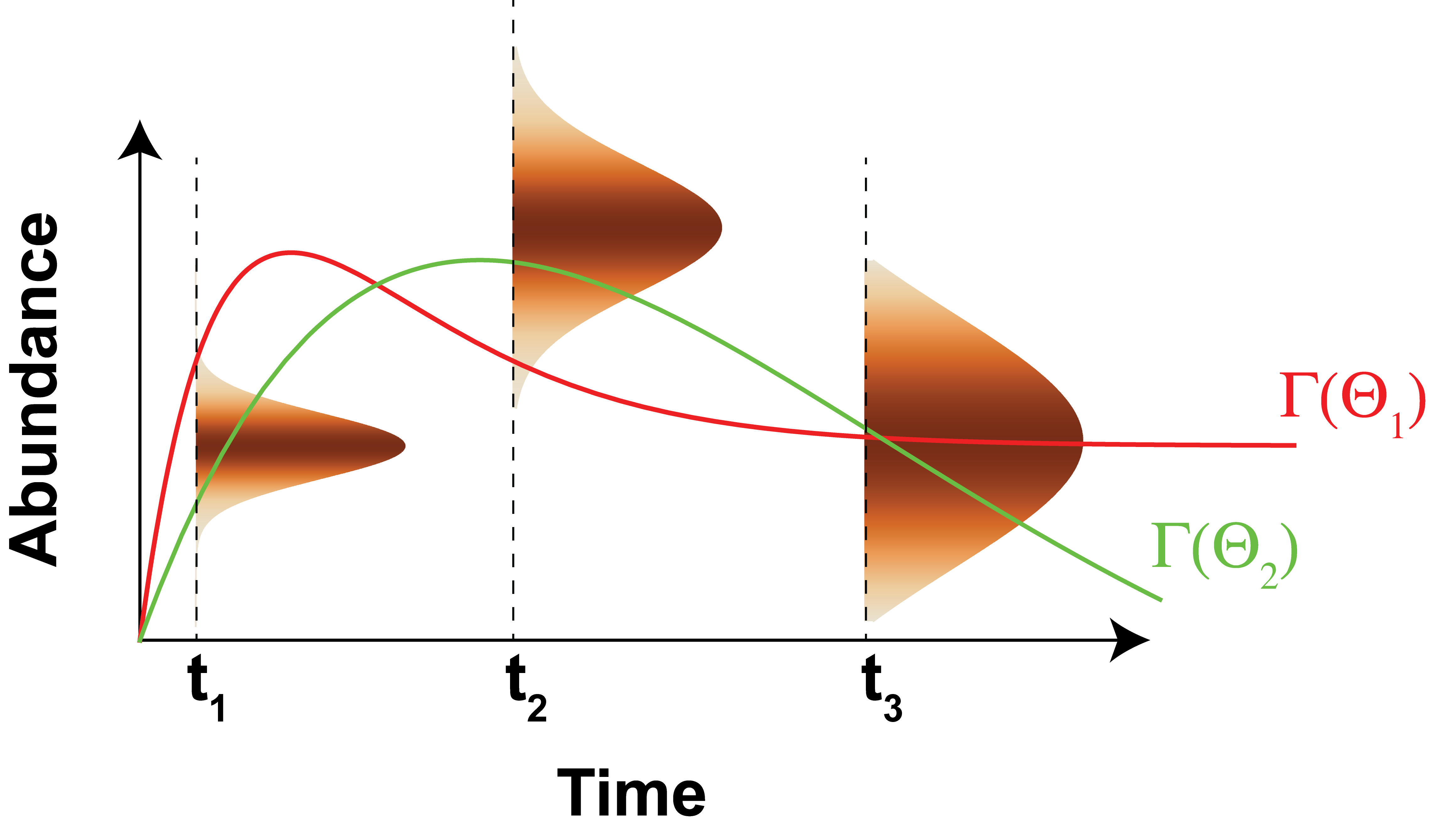}
	\caption{Cell-to-cell variability in the abundance of a chemical species measured at 3 time points $t_1$, $t_2$, and $t_3$. From the experimental data, we determine the fraction $\phi_{ik}$ of cells that populate the $k^{\rm th}$ abundance bin at the $i^{\rm th}$ time point by binning the cell-to-cell variability data in $B_i$ bins. The signaling network model can predict trajectories of species abundances as a function of network parameters (see $\Gamma(\Theta_1)$, $\Gamma(\Theta_2)$, and $\Gamma(\Theta_3)$). Dixit et al.~\citep{dixit2013quantifying,dixit2017maximum} derived the analytical expression for the parameter distribution $P(\Theta)$ as well as the distribution over trajectories $P(\Gamma(\Theta))$ consistent with experimentally estimated populations.\label{fg:ll_bayesian}}
\end{figure}

 More complex biochemical networks can also be treated by Max Cal.  For many biochemical networks (for example bacterial chemotaxis~\citep{wadhams2004making}, mammalian growth factor~\citep{herbst2004review}), the network structure is known, but the individual rate parameters are not.  And, the challenge is to infer them because data collected on only a few species at a few experimental time points and the parameters themselves vary substantially from cell to cell in a population (called \emph{extrinsic variability}).  Dixit et al.~\citep{dixit2013quantifying,dixit2017maximum} have developed a Max-Cal-based framework to infer probability distributions over network parameters and species abundance trajectories in biochemical networks from experimental data.  They inferred the distribution $P(\Theta)$ of network parameters $\Theta$ of a biochemical network as well as the distribution $P[\Gamma(\Theta)]$ over trajectories of species abundances $\Gamma(\Theta)$ that are consistent with histograms of experimentally measured cell-to-cell variability, for example, by flow cytometry or immunofluorescence (see Fig.~\ref{fg:ll_bayesian}).  They showed that the framework can be used to quantify extrinsic noise in both stochastic gene expression networks~\citep{dixit2013quantifying} as well as in signaling networks~\citep{,dixit2017maximum}.
 
 Some probability distributions are not exponential, and have power-law tails, particularly in social and economic systems.  They include distributions of incomes, wealth, city sizes, journal citations, terrorist attacks, and protein-protein interactions, and others.  An entropy variational principle can be applied in some of these cases too, but they use energy-like cost functions that are non-extensive~\citep{GeorgeDillPNAS,peterson2013maximum}.

\section{The foundations of MaxEnt and MaxCal: model-making and physics}

 What is the justification for Max Cal?  Why should populations of fluxes be computable by maximizing a path entropy, subject to a few dynamical constraints?  And, when might it fail?  We view the principle of Max Cal as having the same foundation and justification as other entropy variational principles, but simply applied to pathways, rather than states.  Below, we divide the history of different justifications for this principle into 3 eras: (1) the Boltzmann-Gibbs idea (starting, late 1800's -- )~\cite{landau1968statistical,chandler1987} that statistics and probabilities can be leveraged to compute macro thermo from the micro of the mechanics of particle collisions, then (2) the Jaynes-Shannon idea (starting around 1950's -- )~\cite{shannon,Jaynes57} that entropy maximization is an informational procedure in which you aim to `minimize your ignorance with respect to all except what the data tells you explicitly', then (3) Shore-Johnson (1980 -- )~\cite{ShoreJohnsonIEEE1980} and its interpretation ~\cite{PresseRMP2013} that entropy variation is about making self-consistent models that satisfy the laws of probability.

\subsection{Boltzmann and the mechanics of collisions}  

How can you predict the macroscopic observables of gases and materials from the molecular properties of materials?  Boltzmann and others sought answers in the mechanics and collisions in gases.  Today's molecular simulations typically apply Newton's equations of motion, even for complex systems like protein molecules in water, and then collect statistical properties.  However, Boltzmann's great insight was that, while he could calculate, in principle, all the properties of gases, collision by collision, fully mechanically, he realized that's rarely ever practical or useful.

Rather, his view was that one could capture all the micro-details of the collisions, without computing them, by thinking of them in probabilistic terms instead. This is sufficient to describe equilibrium and evolution of average properties. This key insight was captured by a single relation, $S = k \ln W$, that harnesses `statistics' as a practical approximation for computing the `mechanics'.  This same idea is expressed equivalently as $S = -\sum p_i \ln p_i$ that holds in all different ensembles.  $S$, on the left, is related to the macroscopic property (heat). $W$, on the right, is related to the microscopic details yielding number of arrangements/multiplicity leading to probabilities and distributions.  The right-hand-side is where the statistical interpretation or approximation of mechanics gets embodied. The logarithm can be intuitively understood assuming entropy (like energy) being extensive should add, while multiplicities -- like probabilities -- should multiple. This can be then satisfied by $S = k \ln W$. The idea that free energies are related to populations through $\Delta F_i = -kT \ln p_i$, where $p_i$ is the population of state i, derives from Boltzmann's $S = k \ln W$ combined with the basic thermodynamic relation $F = U - TS$.  This joining together of statistics with mechanics is the basis for practical uses of statistical mechanics today.

However, as noted in a more extensive history elsewhere~\cite{PresseRMP2013}, Boltzmann's arguments -- which converted time-dependent mechanical trajectories to static probability distribution functions -- required the assumption of the \emph{ergodic hypothesis}.  However, even as early as the 1890's, ergodicity was argued to be problematic, by Loschmidt, Zermelo, Poincare and others.  As a modern example of the problem, if you simulate a protein molecule in water (even around its native state), it is usually essentially impossible to sample configurations sufficiently to be certain of seeing all states in proportion to their populations~\cite{SawleGhoshJCTC2016}.  Lack of convergence is just another term for lack of ergodicity. Gibbs' method, described below, avoided the ergodicity premise.

\subsection{Gibbs and the ensemble method}

 J. Willard Gibbs generalized Boltzmann's work so that it could apply to systems of interacting particles. Gibbs devised the logic of \emph{ensembles} and computed probabilities using multinomial statistics, essentially envisioning configurations of a system as samples of a distribution, just like in a dice problem.  There are two issues with Gibbs's method. First, the ensemble is an artificial construct to motivate probabilities using a frequentist interpretation. Second, his predictions were contingent upon an axiom of \emph{equal a priori probabilities} (EAP) over all states of the system regardless of their energy.

\subsection{Shannon's information entropy and Jaynes' \emph{predictive statistical mechanics}} 

 In 1957, Jaynes expressed the view that maximization of the entropy function, subject to a first-moment constraint (on the average energy $\langle E \rangle$), embodies the idea of being maximally ignorant about all the details of the distribution except for that which is needed to satisfy the constraint $\langle E \rangle = k_B T$~\cite{Jaynes57}.  Jaynes' view was a major shift from thinking about statistical mechanics as a physical theory to thinking about statistical mechanics as \emph{information-theoretic}.  On the one hand, this solved key problems -- of no longer requiring ergodicity or the EAP. Jaynes' derivation of the Boltzmann distribution law is also very attractive for didactics; it is simple to teach. Furthermore, it no longer relied on the frequentist argument. However, there were objections to Jaynes' view too.  If entropy is about ignorance, then whose ignorance?  And, why is ignorance relevant at all for problems in physical sciences where entropy can be measured.

\subsection{The Shore-Johnson case about entropy variation} 

  In 1980, Shore and Johnson (SJ) proposed a different, and axiomatic, argument.  They showed that maximizing the Boltzmann-Gibbs entropy is the only procedure that draws inferences about probability distributions that are consistent with the basic rules of probability. SJ interpreted statistical physics~\cite{PresseRMP2013} as the procedure that maximizes $S = -\sum p_i \ln p_i$ subject to constraints when presented with some sort of a model of physical reality that has unknown parameters or features, and the goal is to learn from given data to ultimately infer the posterior distribution that obeys the laws of probability.  It asserts that entropy maximization is the only self-consistent logical pipeline from premises (model, data and a prior distribution) to conclusions (a posterior distribution).

Shore and Johnson established entropy variational principles on a strong footing by showing it is all about self-consistent inferences about models, not about ignorance.  An important value of the inference-based view is that it gives insight into why entropy variation is such a universal idea beyond just material equilibria.  This allows application of statistical mechanical ideas in diverse areas well outside of the scope of material physics for example, in ecology~\citep{phillips2006maximum}, sociology~\citep{peterson2013maximum}, and biology~\citep{BialekNature}.  And, it is not limited to equilibria.  It follows that paths are legitimate objects of probabilities, over which entropies can be maximized, in order to infer dynamics.  This is the basis for Max Cal.

 \section{The issues, caveats and challenges with Maximum Caliber}
 
 It follows from Shore and Johnson that Max Ent or Max Cal are general principles for making self-consistent probabilistic models, not principles of physics, \emph{per se}.  If a model is being used that does not accurately represent the physical situation being treated, then inferences from that model can be wrong.  In such cases, what is flawed is not the logical pipeline of entropy maximization, but rather the premises that are put into that pipeline~\citep{dewar2009maximum}.  This can be either the ignorance about the prior or the lack of data used as constraints. Such instances can be ways to learn from, and improve, models of real world situations by gathering more data.
 
 \subsection{Accounting for measurement errors in the constraints}  

One persistent question is whether entropy maximization can incorporate errors in measurements.  In short, both Max Ent and Max Cal can include them.  First, we discuss how the question arises.  Within the Bayesian framework, entropy maximization can be interpreted as a maximum likelihood problem~\citep{PresseRMP2013}.  Suppose we want to infer a distribution $p(\Gamma)$ over trajectories $\Gamma$ of a system.  Suppose we are constraining an average $\bar f$ of a path-based quantity $f(\Gamma)$.  Now, suppose that there are no errors in the estimate of $\bar f$; we know that value precisely.  Then, we can recast the Max Cal problem as the following maximum likelihood problem
\begin{eqnarray}
{\rm maximize}~e^{\alpha S[p(\Gamma)]} \times \delta \left ( \sum_\Gamma p(\Gamma)f(\Gamma) - \bar f\right ).\label{eq:bayesian}
\end{eqnarray}
In Eq.~\ref{eq:bayesian}, the first term $e^{\alpha S[p(\Gamma)]}$ is the \emph{entropic prior}~\citep{caticha2004maximum,PresseRMP2013} distribution that weighs different candidate distributions $p(\Gamma)$ according to their entropy. The second term is the likelihood. The Dirac delta function here enforces the idea that the estimate of $\bar f$ has no errors or variance associated with it.  Now on the other hand, consider a situation in which we know that there is uncertainty and it is represented by a standard deviation $\sigma_f$ associated with $\bar f$.  In this case, the maximum likelihood problem now becomes:
\begin{eqnarray}
{\rm maximize}~e^{\alpha S[p(\Gamma)]} \times e^{-\frac{\left (\sum_\Gamma p(\Gamma)f(\Gamma) - \bar f \right )^2}{2\sigma_f^2}}.
\label{eq:bayesian_error}
\end{eqnarray}
In Eq.~\ref{eq:bayesian_error}, a Gaussian error distribution is taken into account. Moreover, this Bayesian viewpoint not only allows us to find the distribution $p(\Gamma)$ having the maximum likelihood, but also allows us to express the full posterior distribution $\phi[p(\Gamma)]$ over the distributions $p(\Gamma)$. We have
\begin{eqnarray}
\phi[p(\Gamma)] \propto e^{\alpha S[p(\Gamma)]-\frac{\left (\sum_\Gamma p(\Gamma)f(\Gamma) - \bar f \right )^2}{2\sigma_f^2}}
\end{eqnarray}
 
 \subsection{What constraints are appropriate?  Some situations entail size dependence and some don't}
 
What are the requirements for choosing constraints for Max Ent and Max Cal?  In some classes of problems, only first moments are used.  For example, for macroscopic equilibria, the canonical ensemble is predicted by using a first-moment constraint, namely the average energy $\langle E \rangle$.  In those cases, higher moments are not imposed.  For other situations, higher moments are appropriate (such as modeling gene networks).  How can we rationalize these differences?

At the most basic level, Shore and Johnson showed that any constraint is suitable for Max Ent/Max Cal inference that is linear in the probabilities.  But, this is not very restrictive.  A more precise further division is useful, into:  \emph{scalable} (\emph{size-dependent}) systems, examples of which include equilibria or kinetics of material systems composed of atoms or molecules), vs. \emph{non-scalable} problems, where system size is not a relevant concept (typically problems of data inference or model-making). 

Nonscalable problems are typically inference or model-making situations: we start with some prior distribution function, we then learn some data, and we  want to infer the new posterior distribution. In these nonscalable inference problems, the systems size is not a relevant variable. In nonscalable problems, we are free to use whatever experimental information we have to make inference. In such cases, we can use first or higher moments, or other knowledge~\citep{BialekNature,mora2010maximum,SteveToggle,dixit2017maximum,FirmanBJ2017}. If constraints beyond first moment are negligible it will be seen from the data which will make Lagrange multipliers vanishing for these higher moments. However we cannot assume that apriori. 

 In scalable problems, however, we have the extra predictive power to impose the first moment and ignore all the higher moments at the onset of model building. This is because of
 a particular type of decomposibility into subsystems.  Consider a glass of water as the combination of two half glasses of water, each half-glass having identical intensive properties $(T, p, \mu)$ but each half-glass having half the value of the extensive properties $(E, V, N)$.  Extensive properties scale in proportion to the system size.  Taken to the limit of divisibility, scalable systems can be made up of elemental atoms or molecules or agents or elemental units of some kind~\citep{Dill03}.

 What extra power does this provide?  First, entropy extensivity is the basis for some celebrated results of thermodynamics -- the ability to define equilibria between subsystems by equalities of the intensive variables, the Maxwell relationships, and others.  Second, of interest here,  extensivity defines what constraints are appropriate for Max Ent predictions of large systems.  Consider bringing two subsystems $a$ and $b$ together in a way that allows the exchange of an extensive property, such as the energy $E$, resulting in a combined system.  In typical bulk equilibria, the subsystems are large enough to have precise average values of $\langle E \rangle$, which obey conservation in the exchange, $\langle E \rangle_{\mathrm{total}} = \langle E \rangle_a +  \langle E \rangle_b$\footnote{We limit consideration here to only those where the range of interactions is smaller than the system size.}  In such situations fluctuations are negligible.  Consequently, higher moments are not relevant because they scale sub-linearly with system size, so they become unimportant compared to $\langle E \rangle$.  Similarly, vector properties -- such as the momentum $\langle mv \rangle$ -- are not relevant either, because they vector-average to zero for large systems.  In contrast, for sufficiently small systems, or where scalability is otherwise not applicable, higher-moment constraints often result in more accurate models compared to the traditional statistical mechanical ensembles~\citep{ChamberlinHill,ChamberlinWolfEPJB2009,ChamberlinEPJB2009,dixit2013maximum,dixitPCCP2015,dixit2017mini, dauxois2002dynamics}.
 
 Similarly, first-moment constraints are appropriate to model  dynamics and Max Cal for systems where flows can be regarded as sums of component flows and large enough that average fluxes are well defined, and where $\langle J \rangle_{\mathrm{total}} = \langle J \rangle_a +  \langle J \rangle_b$.  This is the basis for the results above with Green-Kubo, Onsager reciprocal relations, and minimum entropy production~\citep{HazoglouJCP2015}.  Likewise, first moment in number of transitions between discrete states yield Markov processes and Master equation. Additional scalable constraints are sometimes applied, such as the condition of detailed balance, for kinetics that occurs at equilibrium. 
 
However, note that while scalability allows us to discard second moment constraints, and simplify, it does not necessarily guarantee that first moment constraints are sufficient to describe system dynamics. Beyond the first moment constraints used, there may be additional scalable quantities that are relevant. For example if grand canonical ensemble is described using only energy as a constraint, we will get erroneous result because of ignoring the other extensive quantity of particle number. Going a step further, even after knowing all the relevant extensive quantities and using their first moment as a constraint we may be at error, if there are other non-scalable observables at play. Not enough is yet known about what additional constraints are appropriate for far-from-equilibrium dissipative situations~\citep{jack2016,maes2016non}.  As a result, our current lack of complete understanding in how to establish appropriate constraints in those cases should not be taken to imply the failure of the entropy variational principle itself, Max Ent or Max Cal.  Rather, it indicates the need for more experience with more complex dissipative systems and the construction of constraints derived from likelihoods directly motivated from experiments~\citep{PresseRMP2013}.

 
 \section{Summary}
 
 We have discussed Maximum Caliber, a path-entropy-maximization principle for inferring dynamical distributions.  It is quite general -- applicable both near and far from equilibrium, and not limited to material systems in contact with baths.  We show that it recovers, as a general principle should, well-known results of near-equilibrium dynamics -- including the Green-Kubo fluctuation-dissipation relations, Onsager's reciprocal relations, and Prigogine's Minimum Entropy Production.  We describe examples of path-entropy variation results in inferring trajectory distributions from limited data, finding reaction coordinates in bio-molecular simulations, and modeling the complex dynamics of non-thermal systems such as gene regulatory networks or neuronal firing.

\begin{acknowledgments}
KD appreciates support from the National Science Foundation (grant number 1205881) and from the Laufer Center.  SP acknowledges the support for an ARO grant from the Mechanical Sciences Division (66548-EG for Complex Dynamics and Systems), KG acknowledges support from the National Science Foundation (grant number 1149992), Research Corporation for Science Advancement and PROF grant from the University of Denver.
\end{acknowledgments}

%

\end{document}